\documentclass[aps,prl,twocolumn,english,balance,superscriptaddress,floats,showpacs,longbibliography]{revtex4-2}
\usepackage[T1]{fontenc}
\usepackage[latin9]{inputenc}
\usepackage{amsmath}
\usepackage{amssymb}
\usepackage{stmaryrd}
\usepackage{graphicx}
\usepackage{esint}
\usepackage{subfigure}
\usepackage{multirow}
\usepackage{wasysym}
\usepackage{xcolor}
\usepackage{mathtools}
\usepackage{gensymb}
\usepackage[normalem]{ulem}
\usepackage{notes2bib}
\usepackage{bm}
\usepackage{booktabs}
\usepackage{tabularx}
\bibnotesetup{
  note-name = {},       
  use-sort-key = false  
}

\makeatletter

\newcommand{\beq}{\begin{equation}}
\newcommand{\eeq}{\end{equation}}
\newcommand{\bea}{\begin{eqnarray}}
\newcommand{\eea}{\end{eqnarray}}
\newcommand{\bwt}{\begin{widetext}}
\newcommand{\ewt}{\end{widetext}}
\@ifundefined{textcolor}{}
{%
 \definecolor{BLACK}{gray}{0}
 \definecolor{WHITE}{gray}{1}
 \definecolor{RED}{rgb}{1,0,0}
 \definecolor{GREEN}{rgb}{0,1,0}
 \definecolor{BLUE}{rgb}{0,0,1}
 \definecolor{CYAN}{cmyk}{1,0,0,0}
 \definecolor{MAGENTA}{cmyk}{0,1,0,0}
 \definecolor{YELLOW}{cmyk}{0,0,1,0}
}

\newcommand{\bh}{\mathbf{h}}

\newcommand{\add}[1]{{\textcolor{black}{#1}}}

\newcommand{\bj}{\mathbf{j}}

\newcommand{\bE}{\mathbf{E}}
\newcommand{\bP}{\mathbf{P}}
\newcommand{\bd}{\mathbf{d}}

\newcommand{\bk}{\mathbf{k}}

\newcommand{\bq}{\mathbf{q}}
\newcommand{\bp}{\mathbf{p}}
\newcommand{\br}{\mathbf{r}}

\usepackage{babel}
\makeatother
\usepackage{babel}
\begin{document}

\title{Giant nonlinear conductivity in 2D electron gas \add{and single-layer graphene} from substrate-induced dipolar scattering}

\author{Dmitry V. Chichinadze}
\email{cdmitry@wustl.edu}
\author{Alexander Seidel}
\author{Zohar Nussinov}
\affiliation{Department of Physics, Washington University in St. Louis, St. Louis, MO, 63130 USA}

\begin{abstract}
     Despite a surge of interest in the nonlinear transport in 2D materials, a fundamental puzzle remains: existing theoretical frameworks are unable to quantitatively account for the giant nonlinear conductivities ($\gtrsim 1 \frac{\mu \text{m}}{\Omega \text{V}}$) recently reported in 2D van der Waals heterostructures. Here, we introduce a mechanism based on electron scattering from a substrate-induced dipole array linked to short-range impurities. We show that the strict kinematic constraints, inherent to 2D scattering, lead to a singular enhancement of the nonlinear response, fundamentally dictating a natural scale of $1 \frac{\mu \text{m}}{\Omega \text{V}}$. \add{We confirm our estimates by calculating nonlinear conductivity for realistic 2D electron gases and single-layer graphene.}  
\end{abstract}

\maketitle

\textbf{Introduction.}
In recent years, second order nonlinear electron transport in 2D systems with broken inversion symmetry has become a focus of intense research due to new physical effects and potential practical applications \cite{GenkinMednis1968,Onoda2006,Gao2014PRLNHEmetric,Sodemann_Fu,Morimoto2018,KoenigDzeroLevchenkoPesin,IsobeSciAdv2020,Wang2021PRL,Du2021NatComm,Du2021Review,Ortix2021Review,Freimuth2022,AgarwalQG2023PRB,AtenciaNHE2023,SuarezRodriguezetal2025}. In this transport regime, Ohm's law 
acquires quadratic corrections and, in  
component form, reads
\begin{equation}
    j_{i} = \sigma_{ij} E_{j} + \tilde{\sigma}_{ijk} E_{j} E_{k},
\end{equation}
where $\sigma_{ij}$ is a rank-2 linear conductivity tensor and $\tilde{\sigma}_{ijk}$ is a rank-3 nonlinear conductivity tensor. In the DC limit, which we consider in this work, $\tilde{\sigma}_{ijk}$ is symmetric 
under the interchange $j \leftrightarrow k$.

A central and pressing mystery is the origin of the giant second order nonlinear conductivity ranging from $\sim 1 \frac{\mu \text{m}}{\Omega \text{V}}$ to $\sim 10^4 \frac{\mu \text{m}}{\Omega \text{V}}$ \cite{duan2022giantsecondordernonlinearitytwisted,He2022,Chichinadze2025giant,he2025giantfieldtunablenonlinearhall} in 2D metallic systems, as highlighted in, 
e.g., \cite{He2022,Chichinadze2025giant}.
In these works, the nonlinear conductivity tensor extracted from experimental data was found to be orders of magnitude larger than theoretical estimates based on any known mechanism. 
This massive quantitative discrepancy is not merely a numerical detail; it signals a fundamental breakdown in our current understanding of nonlinear transport in 2D materials.

While several candidates have been proposed to bridge this gap, such as quantum corrections to conductivity (which are known to cause nonlinearities in the $I-V$ characteristics of low-dimensional systems \cite{Leadbeater2000,Schwab2001}) in transition metal dichalcogenide van der Waals (vdW) heterostructures \cite{Chichinadze2025WLWAL} and complex electron-electron interaction effects \cite{Morimoto2018}, none have yet provided a universal or quantitatively consistent explanation for the observed scales. In this work, we provide a scenario that resolves the puzzle. We show that the scattering of conduction electrons off a dipole array, induced by ferroelectric or ferrielectric substrates, generates nonlinear conductivity with a natural scale that matches experimental values.
Although nonlinear transport in polar media has a rich history in 3D systems \cite{RustagiPRB1970,Belinicher_1980_Uspekhi,Gorbatsevich1983anomalousJETP,Belinicher1977photogalvanicJETP,Belinicher1980photogalvanic}, we show that in 2D unique scattering anomalies elevate nonlinear response to a giant effect.

The experimental realization of such dipolar environments is common in 2D vdW heterostructures with broken inversion symmetry, where ferroelectric or ferrielectric phases are engineered via external electric displacement fields, interlayer twist angle, or the relative shift between the layers. 
For example, 2D conductors encapsulated by hBN (or other binary compounds) appear to display ferroelectric behavior, as was shown in recent experiments \cite{Pablo_Science_2021,Zhang2024ferroelectricNatComm} and suggested by DFT calculations \cite{Li2017ACS}. In addition to mesoscopic dipole moments, twisted ferroelectric systems were proposed to exhibit electric polarization features similar to merons and anti-merons \cite{BennettNatComm2023,BennettPRR2023}. Importantly, even in bulk ferroelectric systems with polarization vortex textures, the dipoles at surface boundaries exhibit \emph{in-plane} orientation of dipoles \cite{Naumov2004,Yadav2016observation,Damodaran2017,ShaferPNAS2018,Mishra2,Mishra1}. This means that conduction electrons in the immediate proximity of a ferroelectric surface are subject to a robust electric dipole potential.

There is a mounting amount of experimental evidence pointing towards an intimate link between the emergence of large $\tilde{\sigma}$ and the existence of ferro- or ferrielectic behavior in 2D vdW heterostructures \cite{NanoLett2026}. In graphene based heterostructures, for instance, $\tilde{\sigma}$ is dramatically enhanced when graphene is encapsulated between hBN \cite{He2022,he2025giantfieldtunablenonlinearhall}. This configuration is known for its ferroelectric behavior under certain conditions \cite{Pablo_Science_2021}. Similar nonlinear electronic response was observed in other setups and materials intimately linked to ferroelectricity of electron dipole moments present in the system \cite{VladimirMFridkin_1978}.

In this work, we consider a 2D isotropic Fermi gas \add{and single-layer graphene, both} subject to scattering off impurities, which have conventional short range potential and substrate-induced dipole potential with the same magnitude and orientation of dipoles for all impurities. First, by employing symmetry analysis, we show that to linear order in the dipole moment, the $C_3$-symmetric contribution to the nonlinear conductivity tensor vanishes \add{for both isotropic and $C_3$-symmetric dispersions}. Next, we utilize the Boltzmann equation approach \cite{Belinicher1977photogalvanicJETP,Sturman_1984}, calculate the $T-$matrix to linear order in the dipole moment, and solve for distribution functions spanning both linear and nonlinear regimes. Crucially, we obtain that the scattering matrix \add{gradient} is enhanced for forward scattering. 
\add{Crucially, we obtain that the scattering matrix \add{gradient} is enhanced for forward scattering. For a single Fermi surface (the focus of this Letter), energy conservation causes the potentially logarithmically divergent forward scattering contributions to cancel in the momentum space collision integral that determines $\tilde{\sigma}$ in two dimensions. We remark, however, that when scattering connects different bands, the incoming and outgoing momenta may vary differently with energy. Thus, in multiband instances, such a cancellation need not occur. As the Fermi surfaces of different bands approach one another in momentum space,  additional logarithmically enhanced contributions to $\tilde{\sigma}$ may arise from forward scattering to further augment the already giant nonlinear conductivity that we find here for intraband scattering alone.} Similar enhancement of forward scattering  was recently reported for 2D Fermi liquids \cite{Kryhin2023,Kryhin2,NazaryanLevitov}. In our model, however, the enhancement is much more prominent in the nonlinear regime due to the special role of gradients of the distribution functions. Our analysis yields a characteristic scale of $ \tilde{\sigma} \sim 1 \frac{\mu \text{m}}{\Omega \text{V}}$, providing a microscopic mechanism capable of explaining the giant nonlinear conductivity in 2D electron systems.

\textbf{Symmetry analysis.}
The structure of $\tilde{\sigma}_{ijk}$ is determined by the symmetry of a 2D electron gas (2DEG) proximitized to a ferroelectric substrate. We consider a 2DEG situated in the $xy-$plane with the substrate polarization $\bP$ oriented along a fixed, arbitrary axis in 3D. The role of the substrate polarization $\bP$ is to induce dipole moments $\bd$ on initially short-range impurities embedded in 2DEG, see End Matter. These induced dipole moments will be approximately aligned along the polarization axis of the substrate. To permit a non-vanishing DC nonlinear response, the 2D inversion symmetry $(x,y) \rightarrow (-x, -y)$ must be broken. We decompose the coordinate and dipole vectors into in-plane and out-of-plane components: $\br = \br^{\perp} + \br^{\parallel}$ and $\bd = \bd^{\perp} + \bd^{\parallel},$
where $\parallel$ denotes components in the $xy-$plane and $\perp$ indicates the $z-$component. 

For electrons confined to the 2DEG, $\br^{\perp}=z \hat{z} =0$. 
Under \emph{in-plane} inversion, the components transform as $\bd^{\parallel} \rightarrow - \bd^{\parallel}$ and $\bd^{\perp} \rightarrow \bd^{\perp}.$
Since $U^{dip} \propto \br \cdot \bd$, the dipole field will not be experienced by the 2DEG electrons if $\bd \uparrow \uparrow \hat{z}$, hence, the linear $\sigma_{ij}$ and nonlinear $\tilde{\sigma}_{ijk}$ conductivity tensor leading order expansions in $\bd$ read
\begin{equation}
    \begin{gathered}
        \sigma_{ij} E_{j} \simeq \sigma_{ij}(\bd=0) E_{j} + \beta^{d^{\parallel}}_{ijkl} E_{j} d^{\parallel}_{k} d^{\parallel}_{l}, \\
        \tilde{\sigma}_{ijk} E_{j} E_{k} \simeq \tilde{\sigma}_{ijk} (\bd=0) E_{j} E_{k} + \gamma^{d^{\parallel}}_{ijkl} E_{j} E_{k} d^{\parallel}_{l}.
    \end{gathered}
    \label{P_phenom_correct_no_Pz}
\end{equation}
These expressions are consistent with \cite{Gorbatsevich1983anomalousJETP}. 
We note that the linear conductivity acquires only a quadratic correction, whereas the nonlinear conductivity contribution starts from the linear order in $\bd$.

The structure of $\bd^{\parallel}$-induced nonlinear conductivity is further elucidated by a symmetry analysis of the nonlinear Ohm relations. In the complex representation, quadratic corrections to Ohm's law in 2D are expressed as 
\cite{Chichinadze2025giant}
\begin{equation}
\begin{gathered}
    j_x^{(2)} + i j_y^{(2)} =  \Xi^{(2)}_{-} (E_x - i E_y)^2 + \\ + \Xi^{(2)}_{+} (E_x + i E_y)^2 + \Xi^{(2)}_{0} (E_x^2 + E_y^2), \nonumber
\end{gathered}
\end{equation}
where in-plane rotation follows $v_x \pm i v_y \rightarrow e^{\pm i \alpha} \left( v_x \pm i v_y \right)$. By applying an in-plane rotation to $\bj$ and $\bE$ vectors one can see that the $(E_x - i E_y)^2$-term is invariant upon $120^{\circ}$ rotation (3-fold symmetric), while the other two terms are invariant only upon $360^{\circ}$ rotation.
Adding an in-plane dipole moment $d_x + id_y$ up to the linear order in $|\bd|$ leads to 
\begin{equation}
\begin{gathered}
    j_x^{(2)} + i j_y^{(2)} =  \beta_{+} (d_x - id_y) (E_x + i E_y)^2 + \\ + \beta_{0} (d_x + id_y) (E_x^2 + E_y^2), 
    \end{gathered}
    \label{dipole_sym_corr_an_1} 
\end{equation}
where $\beta_{+},\beta_{0}$ are two complex constants, see SI for details. Fully longitudinal nonlinear response is governed by a 2D vector $\mathcal{B}$ which satisfies $\mathcal{B}_x + i \mathcal{B}_y =\Xi^{(2)^*}_{+} + \Xi^{(2)}_{0}$, while nonlinear Hall response (fully transverse) is governed by a vector $\mathcal{C}$, whose components follow $i \mathcal{C}_x + \mathcal{C}_y = \Xi^{(2)}_{+} - \Xi^{(2)^*}_{0}$ \cite{Chichinadze2025giant}. For a circular Fermi surface (FS), without loss of generality, we can pick our reference frame such that $\bd \uparrow \uparrow \hat{x}$. Then, $d_y=0$ and by $M_y$ mirror symmetry (around the $xz-$plane)
$
    \mathrm{Im}\left( \Xi^{(2)}_{+} \right) = \mathrm{Im}\left( \Xi^{(2)}_{0} \right) = 0. \nonumber
$
It follows that $\mathcal{B} \perp \mathcal{C}$, satisfying
\begin{equation}
    \mathcal{B}_y \equiv 0, \;\;\;
    \mathcal{C}_x \equiv 0, 
\end{equation}
i.e., $\mathcal{B} \parallel \bd$ and $\mathcal{C} \perp \bd$.
The signs of the nonzero components $\mathcal{B}_x$ and $\mathcal{C}_y$ are determined by the dispersion-specific constants $\beta_{+},\beta_{0}$. 
Therefore, for a circular FS in the presence of dipoles with moderate $|\bd|$, the 3-fold rotationally symmetric contribution to $\tilde{\sigma}$ is vanishingly small \footnote{\add{As we show in the SI, the 3-fold contribution appears only at order $|\bd|^3$ for both isotropic and 3-fold symmetric electron dispersions.}}, in contrast to the purely longitudinal $\mathcal{B}$ and nonlinear Hall $\mathcal{C}$ contributions.

\begin{figure}[h!]
\includegraphics[width=0.9\linewidth]{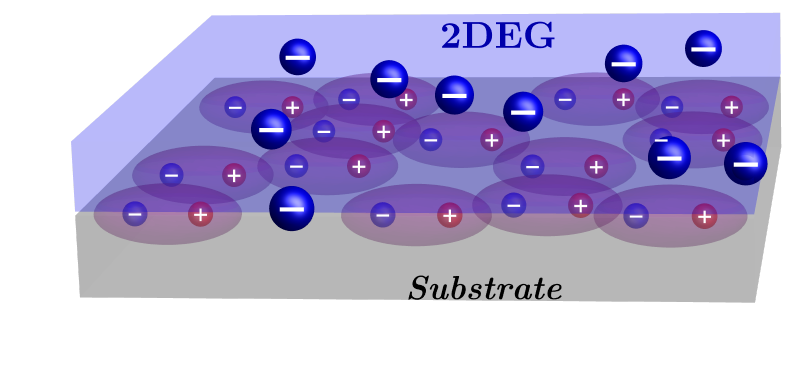}
\caption{Schematic of the physical system. Electrons in a 2D gas (blue) scatter from an array of electric dipoles (purple) induced by the substrate (gray). Nonlinear electric susceptibility of the ferroelectric or ferrielectric substrate stimulates formation of induced dipole moments on short-range impurities, embedded in 2DEG.  }
\label{fig1}
\end{figure}

\textbf{The model.} We consider a 2D electron system scattering of an array of impurities with dipole potential induced on every impurity by a proximity to a ferroelectric or ferrielectric substrate, see End Matter for details.
The disorder potential is given by $U (\br) = \sum_{\br_i} \left[ U^{s} (\br - \br_i) + U^{dip} (\br - \br_i) \right]$, where $U^{s} (\br) = a \delta (\br)$ is the short-range potential and $U^{dip} (\br)$ is the substrate-induced dipole potential.   The Fourier transform of the potential from a single dipole is (see SI)
\begin{eqnarray}
    \add{U^{dip}_{\bk}}  &=& \frac{- 2 \pi i e \bd \cdot  \bk}{\add{\varepsilon_r} k},
    \label{dip_pot_main}
\end{eqnarray}
where $\bd$ is the dipole moment of a single dipole, \add{$\varepsilon_r$} is the dielectric permittivity, and all dipoles are assumed to have the same moment and orientation.  
The setup of the scattering problem is analogous to that in \cite{Belinicher1977photogalvanicJETP,Belinicher1980photogalvanic}, except the system is two-dimensional and inversion is a 2D symmetry.

To calculate $\sigma$ and $\tilde{\sigma}$, we employ the Boltzmann equation \cite{Sturman_1984} \footnote{\textcolor{black}{The two collision integrals $- \int \frac{d^2 \bk'}{(2 \pi)^2} W_{\bk \bk'} \left( f(\bk) - f(\bk') \right)$ and $\int \frac{d^2 \bk'}{(2 \pi)^2} \left( W_{\bk\bk'} f(\bk') - W_{\bk'\bk} f(\bk)  \right)$  are identical if the optical theorem is satisfied, which is true in our model setup.} }
\begin{eqnarray}
    -\frac{e}{\hbar} \bE \frac{\partial f(\bk)}{\partial \bk} = - \int \frac{d^2 \bk'}{(2 \pi)^2} W_{\bk \bk'} \left( f(\bk) - f(\bk') \right),
\end{eqnarray}
where the scattering probability (matrix) $W_{\bk \bk'}$ satisfies
\begin{equation}
    W_{\bk \bk'} = \frac{2\pi}{\hbar} \left| T_{\bk, \bk'} \right|^2 \delta \left( \varepsilon_{\bk} - \varepsilon_{\bk'} \right),
\end{equation}
with scattering being controlled by a $T-$matrix $T_{\bk, \bk'}$.
To study the effect of dipole scatterers on electron transport in the system we follow \cite{Belinicher1977photogalvanicJETP,Belinicher1980photogalvanic} and calculate the $T-$matrix up to the leading term in $\bd$ \add{for an inversion-preserving 2-band Hamiltonian
    $H (\bp) = h_0 \tau_0 + \bh (\bp) \cdot \bm{\tau},$
with $h_z=0$ to eliminate effects of conventional Berry phase skew scattering and side jumps. Above, $\bm{\tau}$ -- is the vector of Pauli matrices in sublattice space.  The expression \add{for band $\zeta = \pm 1$} reads
\begin{widetext}
\begin{equation}
    \begin{gathered}
        \left| T_{\bk, \bk'} \right|^2 = n_{imp} a^2 \left| \langle u_{\bk, \zeta} | u_{\bk', \zeta} \rangle \right|^2 + n_{imp} \frac{4 a^2 \pi^2 e }{\varepsilon_r} \sum_{\bp} z_{\bk' \bk \bp} \; \bd \cdot \left[ \frac{\bk - \bk'}{| \bk - \bk' |}  - \frac{\bk - \bp}{| \bk - \bp |} - \frac{\bp - \bk'}{| \bp - \bk' |} \right]  \delta (\omega - \varepsilon_{\bp} ) \Biggr|_{\omega = \varepsilon_{\bk}},
    \end{gathered}
    \label{Tmodsquared}
\end{equation}
where $n_{imp}$ is the impurity concentration, and following \cite{KoenigDzeroLevchenkoPesin}, we introduced $\hat{h}_{\bk} = \frac{{\bh}_{\bk}}{|{\bh}_{\bk}|}$,  $\left| \langle u_{\bk, \zeta} | u_{\bk', \zeta} \rangle \right|^2 = \frac{1+\hat{h}_{\bk} \cdot \hat{h}_{\bk'}}{2}$ and 
\begin{eqnarray}
    z_{\bk \bk' \bk''} = \langle u_{\bk, \zeta} | u_{\bk', \zeta} \rangle \langle u_{\bk', \zeta} | u_{\bk'', \zeta} \rangle \langle u_{\bk'', \zeta} | u_{\bk, \zeta} \rangle = \left( 1 + \hat{h}_{\bk} \cdot \hat{h}_{\bk'} + \hat{h}_{\bk'} \cdot \hat{h}_{\bk''} + \hat{h}_{\bk''} \cdot \hat{h}_{\bk}  \right)/4.
\end{eqnarray} 
\end{widetext}
}

We then introduce the standard parameterization for distribution functions at linear $f_1 (\bk) = \Phi_1 (\bk) \frac{\partial f_0 (\varepsilon)}{\partial \varepsilon} \biggr|_{\varepsilon = \varepsilon (\bk)}$ and quadratic $f_2 (\bk) = \Phi_{2,1} (\bk) \frac{\partial f_0 (\omega)}{\partial \omega} \biggr |_{\omega=\varepsilon_{\bk}} + \Phi_{2,2} (\bk) \frac{\partial^2 f_0 (\omega)}{\partial \omega^2} \biggr |_{\omega=\varepsilon_{\bk}}$ order in electric field, where $f_0 (\varepsilon_{\bk}) = n_F(\varepsilon_{\bk})$ -- is the equilibrium distribution function at the Fermi surface. After such parameterization and introducing scattering rate $g(\bk) = \int \frac{d^2 \bk'}{(2 \pi)^2} W_{\bk \bk'}$, the three equations that define linear and nonlinear electron transport properties are \cite{Chichinadze2025giant}  
\begin{align}
    &&\Phi_1(\bk) = \frac{e}{\hbar} \frac{\bE}{g(\bk)} \cdot \frac{\partial \varepsilon_{\bk}}{\partial \bk} + \int \frac{d^2 \bk'}{( 2\pi)^2} \frac{W_{\bk, \bk'}}{g(\bk)} \Phi_1 (\bk'), \label{Phi1eq} \\
    &&\Phi_{2,1} (\bk)   = \frac{e \bE}{\hbar g(\bk)} \cdot \frac{\partial \Phi_1 (\bk)}{\partial \bk} +  \int \frac{d^2 \bk'}{(2 \pi)^2} \frac{W_{\bk,\bk'}}{g(\bk)}  \Phi_{2,1} (\bk'), \label{Phi21eq} \\
    &&\Phi_{2,2} (\bk)   = \frac{e \bE \cdot \frac{\partial \varepsilon(\bk)}{\partial \bk}}{\hbar}  \frac{\Phi_1 (\bk)}{g(\bk)} +  \int \frac{d^2 \bk'}{(2 \pi)^2} \frac{W_{\bk,\bk'}}{g(\bk)}  \Phi_{2,2} (\bk'). \label{Phi22eq}
\end{align}
Solving for three $\Phi-$functions allows then to calculate electron current density using
\begin{equation}
    \begin{gathered}
        j_{\alpha} = -\frac{e}{\hbar} \int \frac{d^2 \bk}{(2 \pi)^2}  \frac{\partial \varepsilon(\bk)}{\partial k_{\alpha}} \Phi_1 (\bk) \frac{\partial f_0 (\varepsilon)}{\partial \varepsilon} \biggr|_{\varepsilon = \varepsilon (\bk)} - \\
        - \frac{e}{\hbar} \int \frac{d^2 \bk}{(2 \pi)^2} \left( \frac{\partial \varepsilon(\bk)}{\partial k_{\alpha}} \Phi_{2,1} (\bk)  -  \frac{\partial \Phi_{2,2} (\bk)}{\partial k_{\alpha}} \right) \frac{\partial f_0 (\omega)}{\partial \omega} \biggr |_{\omega=\varepsilon_{\bk}}.
        \label{current_density}
    \end{gathered}
\end{equation}

\textbf{The role of gradients.}
Closed form analytical solutions for the functions $\Phi$ and the conductivities $\sigma, \tilde{\sigma}$ may be obtained within our model, \add{as we show in the End Matter and SI}. \add{However, the physical mechanism of nonlinear conductivity enhancement from dipolar scattering in 2D can be understood already from Eqs. \eqref{Phi21eq}--\eqref{current_density}. At the lowest order beyond the Born approximation, the antisymmetric scattering probability satisfies \cite{Belinicher_1980_Uspekhi}
\begin{equation}
    \begin{gathered}
        \add{W^{as}_{\bk\bk'} \propto \mathrm{Im} \int  d^2 \bk'' U_{\bk- \bk'} U_{\bk' - \bk''} U_{\bk'' - \bk} \delta(\varepsilon_{\bk}-\varepsilon_{\bk'}) \delta(\varepsilon_{\bk'}-\varepsilon_{\bk''}) }, \nonumber
    \end{gathered}
\end{equation}
where the disorder potential contains both short-range and dipolar components; at linear order in $|\bd|$ this expression leads to Eq. \eqref{Tmodsquared}. 
Nonlinear current density depends on \emph{gradients} of $\Phi_1 (\bk)$ and $\Phi_{2,2} (\bk)$.
These gradients, in turn, involve $\frac{\partial W_{\bk \bk'}}{\partial \bk} \propto \frac{\partial U^{dip}_{\bk}}{\partial \bk} \propto \frac{1}{k}$ due to the specific form \eqref{dip_pot_main} of the 3D dipole potential embedded into 2D periodic system. }

\add{The enhancement is also a consequence of kinematic constraints in 2D scattering, similar to the collinear scattering probability enhancement described in \cite{Kryhin2023}. Notably, for the dipole array (because of the form of the potential in Eq. \eqref{dip_pot_main}), the forward scattering probability \emph{gradient} $\frac{\partial W_{\bk, \bk'}}{\partial \bk}$ is enhanced. This leads to a significant enhancement of the gradient terms, making the effect much more prominent in the nonlinear transport regime than in the linear one. }

\add{
\begin{table*}[t]
\centering
\begin{tabular*}{\textwidth}{@{\extracolsep{\fill}} l  c c c c c c c}
\toprule
Reference & $\varepsilon_r$ & $n$, (cm$^{-2}$) & $\mu$, (cm$^{2}$/V\,s)
       & $k_F$,  (\AA$^{-1}$) & $\tau$, (s) & $\sigma$, $\left( \frac{e^2}{\hbar} \right)$ & $\tilde{\sigma}$, $\left( \frac{\mu \text{m}}{\Omega \text{V}} \right)$ \\
\midrule
GaAs, \cite{Sammon2018}
  & 12.9 & $3.0\times10^{11}$ & $3 \times10^{7}$ & $1.37\times10^{-2}$ & $1.14\times10^{-9}$ & $5.9 \times 10^3$ & 178 \\
GaAs, \cite{Rossler2010}
  & 12.9 & $2.0\times10^{11}$ & $1.9\times10^{6}$ & $1.1\times10^{-2}$ & $7.2\times10^{-11}$ & 249 & 0.57 \\
GaAs, \cite{Rossler2010}
  & 12.9 & $2.0\times10^{11}$ & $4.5\times10^{6}$ & $1.1\times10^{-2}$ & $1.7\times10^{-10}$ & 588 & 3.2 \\
ZnO, \cite{Falson2016}
  & 8.5  & $8.3\times10^{10}$ & $1.3\times10^{6}$ & $7.2\times10^{-3}$ & $2.2\times10^{-10}$ & 70 & 5.3 \\
\bottomrule
\end{tabular*}
\caption{ \add{Representative parameters of GaAs/AlGaAs and MgZnO/ZnO two-dimensional
electron gases. The Fermi wave vector is estimated as $k_F=\sqrt{2\pi n}$
(spin/valley degeneracy $g_{s/v}=2$) and the transport scattering time as
$\tau = m^*\mu/e$, where $\mu$ stands for mobility, with $m^*=0.067\,m_e$ for GaAs~\cite{Sammon2018} and
$m^*=0.30\,m_e$ for ZnO \cite{Tabrea2020}. Right column shows $\tilde{\sigma} = |\mathcal{B}|=|\mathcal{C}|$ for $d=1 e \cdot \text{\AA}$. }  }
\label{tab:2DEG_params}
\end{table*}
}

\add{\textbf{Approximate analytical solution for 2DEG and scale of the nonlinear conductivity.}}
\add{We first calculate transport properties for a single-band isotropic 2DEG with dispersion $\varepsilon_{\bk} = \frac{\hbar^2 \bk^2}{2m} - E_F$. }
\add{Although for an isotropic 2DEG with $\left| T_{\bk, \bk'} \right|^2 \propto \bd$ an \emph{exact} mathematical solution can be obtained, as we show in the SI, it is instructive to solve for distribution functions up to the linear in $\bd$ accuracy. To do that, we separate all distribution functions as
    $\Phi_1 (\bk) = \phi_0 (\bk) + \phi_1 (\bk),$
where $\phi_n (\bk) \propto |\bd|^n$. In such case, a simple analytical expression can be obtained for nonlinear conductivity tensor, while linear conductivity is given by the standard Drude result. Within this approximation nonlinear current is given by  
\begin{equation}
    \begin{gathered}
        \bj^{(2)} = \bE \left( \mathcal{B} \cdot \bE \right) +  (\hat{z} \times \bE) (\mathcal{C} \cdot \bE) = \\
        g_{s/v} \frac{e^2}{\hbar}   \left( \frac{e^2 k_F \tau^2}{4\pi \hbar^2 \varepsilon} \right) \left[ \bE \left( \bd \cdot \bE \right) +  (\hat{z} \times \bE) (\left( \hat{z} \times \bd \right) \cdot \bE) \right],
    \end{gathered}
    \label{2DEG_nonlin_sigma}
\end{equation}
in accordance with our symmetry analysis. Here $g_{s/v}$ is the combined spin/valley degeneracy factor; for a single valley spin-degenerate FS $g_{s/v} = 2$. It follows directly from Eq. \eqref{2DEG_nonlin_sigma} that for an isotropic 2DEG $\left| \mathcal{B} \right| = \left| \mathcal{C} \right|$.
}

Giant nonlinear conductivity has been reported in 2D materials with magnitudes ranging from $\sim 1 \frac{\mu \text{m}}{\Omega \text{V}}$ to $\sim 10^4 \frac{\mu \text{m}}{\Omega \text{V}}$ \cite{duan2022giantsecondordernonlinearitytwisted,He2022,Chichinadze2025giant,he2025giantfieldtunablenonlinearhall}. 
\add{To estimate the magnitude of the nonlinear conductivity in our model, we consider realistic examples of 2DEGs based on ZnO/MgZnO and GaAs/AlGaAs. Realistic parameters of 2DEGs and the corresponding calculated $\sigma$ and $\tilde{\sigma}$ are shown in Table \ref{tab:2DEG_params}. In contrast to other proposed mechanisms, the enhancement here is a consequence of the physical scattering anomaly, which fundamentally dictates the $\frac{\mu \text{m}}{\Omega \text{V}}$ scale for various realistic parameter sets.
}



\textbf{\add{Calculation for single layer graphene.}}
\add{To further elucidate the applicability of our results to 2D van der Waals heterostructures, we numerically calculate linear and nonlinear conductivity tensors following the generic approach outlined in the SI Sec. IV. Specifically, we consider the tight-binding model of single layer graphene with only nearest neighbor hopping included, which results in 
\begin{equation}
    H (\bp) = 
    t_0 \begin{pmatrix}
    0 && \sum_{j=1}^3 e^{i \bm{\delta}_j \cdot \bp}  \\
    \sum_{j=1}^3 e^{-i \bm{\delta}_j \cdot \bp} && 0
    \end{pmatrix}, 
    \label{grapheneHtb}
\end{equation}
where $\delta_1 = a_l (1,0), \delta_2 = \frac{a_l}{2}(-1, \sqrt{3}), \delta_3 = \frac{a_l}{2}(-1, -\sqrt{3})$ -- are three bond vectors with $a_l=1.42 \text{\AA}$  and hopping integral is $t_0 = 2.8 \text{eV}$.
 }

\begin{figure*}[t]
\includegraphics[width=1.\linewidth]{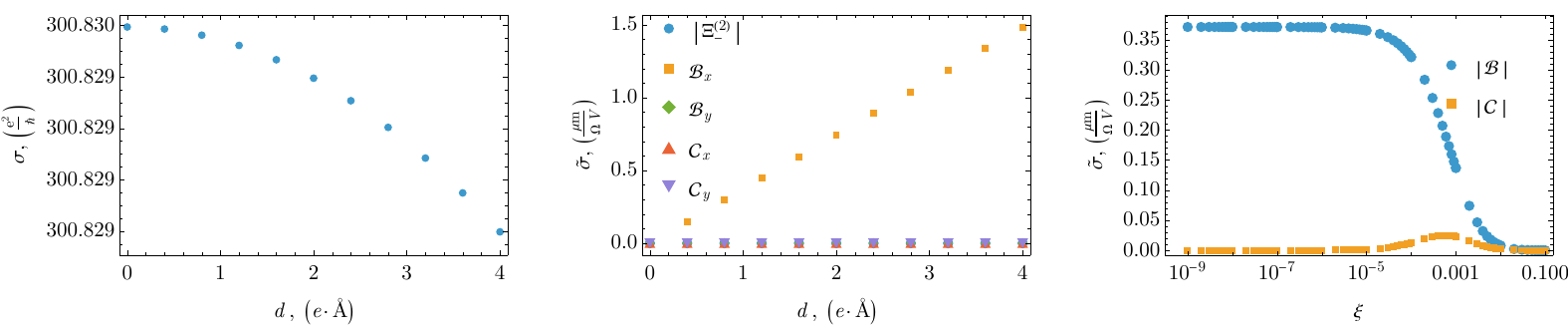}
\caption{ \add{Numerical calculation for a tight-binding model of single-layer graphene, Eq. \eqref{grapheneHtb}, including both spins and both valleys. Numerical solution for angular harmonics obtained using Fast Fourier Transform. For calculations in all panels we added a screening parameter $\xi = k_{scr}/k_F$ to the dipole potential as $U^{dip} ( \bk ) = \frac{- 2 \pi i e \bd \cdot  \bk}{\add{\varepsilon_r} \left( k + \xi \right) }$. (Left) Linear conductivity as a function of dipole moment. (Middle) Components of the nonlinear conductivity tensor as a function of dipole moment. (Right)  Nonzero components of the nonlinear conductivity tensor as a function of $\xi$ (log-linear plot). We used $\varepsilon_r = 3, n_{imp} a^2 = 0.151 \text{eV}^2 \text{\AA}^2, E_F = 8 \times 10^{-3} \text{eV}$ for all panels; $\xi = 10^{-8}$ in the left and middle panels, and $|\bd| = 1 e\mathring{A}$ in the right panel. $\varepsilon_r$ value is chosen to reflect samples with hBN from one side and vacuum from another. } }
\label{figGraphene}
\end{figure*}

\add{The results of our numerical calculations for both spins and both valleys are presented in Fig. \ref{figGraphene}. As is evident from the middle panel, nonlinear conductivity remains $\tilde{\sigma} \sim 1 \frac{\mu \text{m}}{\Omega \text{V}}$ and grows with $|\bd|$.}

\textbf{Discussion and outlook.}
In summary, we have shown that the scattering of 2D electrons from an array of impurities with substrate-induced dipole moments provides a powerful and robust mechanism for generating giant nonlinear conductivity.  Our theory naturally predicts a characteristic scale of $\tilde{\sigma} \sim 1 \frac{\mu \text{m}}{\Omega \text{V}}$, providing a quantitative resolution to recent experimental puzzles.
We also find that to leading order in the dipole moment, the three-fold symmetric contribution to the nonlinear conductivity tensor vanishes -- a result dictated by the symmetry of in-plane dipoles and resembling recent observations \cite{Chichinadze2025giant}.

The microscopic origin of this giant enhancement is a phenomenon specific to 2D, arising from the singular behavior of forward (collinear) scattering. While the setup is reminiscent of classic problems in polar materials \cite{Gorbatsevich1983anomalousJETP,Belinicher1977photogalvanicJETP,Belinicher1980photogalvanic}, the crucial detail here is the 2D geometry, which imposes stricter kinematic constraints for scattering in Fermi liquids and gases \cite{Kryhin2023}. Most importantly, this enhancement is an intrinsic feature of the nonlinear transport regime; it is driven by the scattering probability gradients, $\frac{\partial W_{\bk, \bk'}}{\partial \bk},$ which remain latent in the linear response but emerge as the dominant factor in \emph{nonlinear} current generation.

Notably, this mechanism is quite general and not limited to dipole potentials. Higher order multipoles, such as octupoles, could also yield enhancement of collinear scattering for higher-order in $\bE$ response \footnote{\add{Electrostatic potential for multipoles at $z=0$ can be estimated as 
$V(\br) \propto \frac{1}{r^{n+1}},$
where $n=0$ for monopole, $n=1$ for dipole, $n=2$ for quadrupole, $n=3$ for octupole etc. 
Now, to generate a $1/q$ singularity from this potential we would need to take powers of gradients in $q-$space. We get
$
    \left( \nabla_{\bq} \right)^k V(\bq) \propto \frac{\partial^k}{\partial q^k} \left( q^{n-1} \right) = q^{n-1-k},
$
hence we need $n=k$ to have an $q^{-1}$ singularity. One can show that $k-$th gradient of the potential appears in the integral equation for distribution function $\Phi_{k,l}$ describing response at the order $\bE^k$   \cite{higherMultipoles}.}} \add{  \cite{higherMultipoles}}, potentially realizable with piezoelectric substrates \cite{VladimirMFridkin_1978, Belinicher_1980_Uspekhi}.  The net effect in such systems, however, will depend on a delicate interplay between stronger scattering singularities and typically smaller multipole moments, as large higher-order multipole moments are rather rare in condensed matter systems \cite{Kuramoto2008electronic,Kuramotoetal2009,SantiniRMP2009,Zenin2020engineering}.

By bridging the gap between fundamental results in ferroelectric response theory 
\cite{Belinicher1977photogalvanicJETP} and 2D scattering anomalies \cite{Kryhin2023}, our work establishes a scenario for giant nonlinear conductivity generation. These results not only clarify the nature of nonreciprocal transport in 2D systems but also pave the way for engineering of nonlinear conductivity tensors through substrate control -- a promising direction for 2D electronics and quantum device applications.

\textbf{Acknowledgments.}
We thank E. Henriksen, A. Kamenev, S. Kryhin, A. Levchenko, JIA Li, R. Mishra, D. Shaffer, P. Sukhachov, O. Vafek for useful discussions. 
D.V.C.
acknowledges financial support from Washington University in St. Louis through Edwin Thompson Jaynes Postdoctoral Fellowship. \add{Claude Opus 5.5 by Anthropic was used to find model parameter values for realistic examples. }

\section*{End matter}

\subsection*{Microscopic mechanism of generating dipole moment on short-range impurities}

\begin{figure}[h!]
\includegraphics[width=0.9\linewidth]{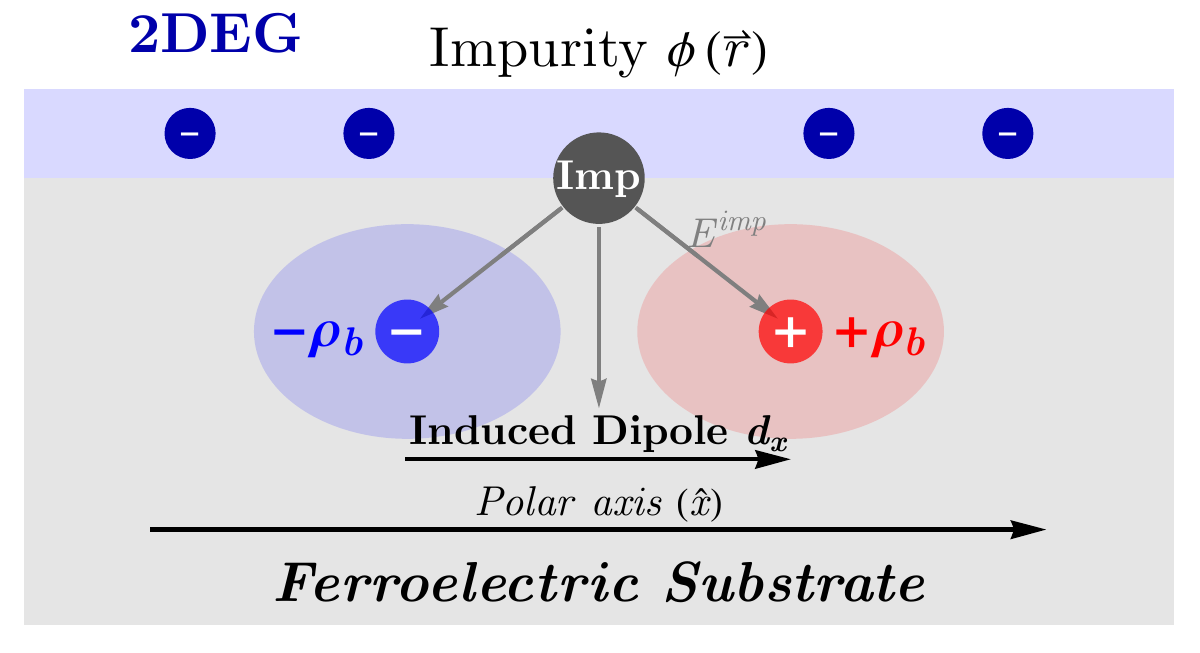}
\caption{ \add{Illustration of the mechanism behind the induced dipole generation on a single impurity. Asymmetric bound charge is generated at the interface between the 2D metal and ferroelectic substrate with a polar axis.  } }
\label{fig_induced_dipole}
\end{figure}

Consider a conducting 2D material proximitized to a ferroelectric substrate with an in-plane polarization. In such case, short-range impurities embedded into a 2D conductor will experience effects due to a specific orientation of electric dipoles on the surface of a ferroelectric substrate, such as anisotropic dielectric screening of impurities and induced dipoles due to bound charge. Here, we will only discuss the second mechanism, although effects of anisotropic screening are of definite interest. 

Short-range impurities create a local perturbation in the electric field $\bE^{imp}$. This field, in turn, modifies the intrinsic polarization of a ferroelectic (or ferrielectric), therefore inducing a slight change in local polarization $\bP_{loc} = \bP + \Delta \bP$. Since the substrate has a preferred orientation of dipoles, the induced correction to polarization will be pointing mainly along the polar axis (due to the specific form of the dipole potential, only the in-plane polarization will have any effect on the conducting electrons in a 2D conductor). Such a change in polarization will create a bound charge density around the impurity: 
\begin{equation}
    \rho_{b} = - \nabla \cdot \Delta \bP,
\end{equation}
thus creating a bound charge of different sign on opposite sides of the impurity.  

To see that explicitly, consider an isotropic short-range electrostatic potential of the impurity $\phi (\br)$, which is symmetric upon inversion: $\phi (\br) = \phi (-\br)$. The impurity electric field 
$\bE^{imp} = - \nabla_{\br} \phi (\br)$ is odd and, e.g., $E_{x}^{imp} (-x,y) = -E_x^{imp} (x,y).$  Let's now assume, without loss of generality, that the substrate polar axis is along $\hat{x}$-axis. Now, assuming strongly anisotropic susceptibility tensor, the total polarization in the ferroelectric substrate along the polar axis is given by
\begin{equation}
    P_x = P_0 + \epsilon_0 \chi^{(1)} E^{imp}_x + \epsilon_0 \chi^{(2)} \left(E_x^{imp}\right)^2 + ... 
\end{equation}
Hence, the induced change in polarization from the impurity field 
\begin{equation}
    \Delta P_x \simeq \epsilon_0 \chi^{(1)} E^{imp}_x + \epsilon_0 \chi^{(2)} \left(E_x^{imp}\right)^2.
\end{equation}
The bound charge, therefore, reads
\begin{equation}
    \begin{gathered}
        \rho_{b} = - \nabla \cdot \Delta \bP = - \frac{\partial \Delta P_x}{\partial x} \simeq \\ \simeq - \epsilon_0 \chi^{(1)}  \frac{\partial E^{imp}_x}{\partial x} - \epsilon_0 \chi^{(2)}  \frac{\partial \left[ \left(E^{imp}_x \right)^2 \right]}{\partial x}.
    \end{gathered}
\end{equation}
The first term is even, as $E_x^{imp}$ is odd, while the second term is odd, since $\left( E_x^{imp} \right)^2$ is even. Thus, there will be an odd contribution to the bound charge $\rho_b,$ which is exactly the induced dipole moment.

The induced dipole moment can be estimated by calculating the dipole moment of the induced bound charge around the impurity: 
\begin{eqnarray}
    d_x = \int d^3 \br \rho_b(\br) x = - \epsilon_0 \chi^{(2)} \int d^3 \br  \frac{\partial \left[ \left(E^{imp}_x \right)^2 \right]}{\partial x} x, 
\end{eqnarray}
as the linear in $E_x^{imp}$ contribution vanishes because an even function is integrated with an odd function. $\bE^{imp}$ is strongly localized due to the short-range nature of impurity potential, therefore, $E_x^{imp}$ decays rapidly. This leads to the vanishing of the boundary term in the integral 
\begin{equation}
    \int_{-\infty}^{\infty} \frac{\partial \left[ \left(E^{imp}_x \right)^2 \right]}{\partial x} x dx = \left[ x \left(E^{imp}_x \right)^2 \right]_{-\infty}^{\infty} - \int_{-\infty}^{\infty} \left(E^{imp}_x \right)^2 dx
\end{equation}
and the induced dipole moment on the impurity reads
\begin{equation}
    d_x = \epsilon_0 \chi^{(2)} \int d^3 \br \left(E^{imp}_x \right)^2.
\end{equation}

The case of ferrielectic substrates or substrates with polar axis but randomly positioned dipoles will be qualitatively the same but differ quantitatively, specifically due to the possible necessity to average over the induced impurity dipole moment magnitude due to random distribution of substrate dipoles in the latter case. \add{We also note, that substrates without a well-defined polar axis would lead to the distribution of individual dipoles' directions, which would require averaging over the directional distribution. Such phenomenon would be highly dependent on the angular distribution of dipoles' directions and will contribute to the supperssion of nonlinear conductivity. }

\begin{figure}[h]
\includegraphics[width=1.0\linewidth]{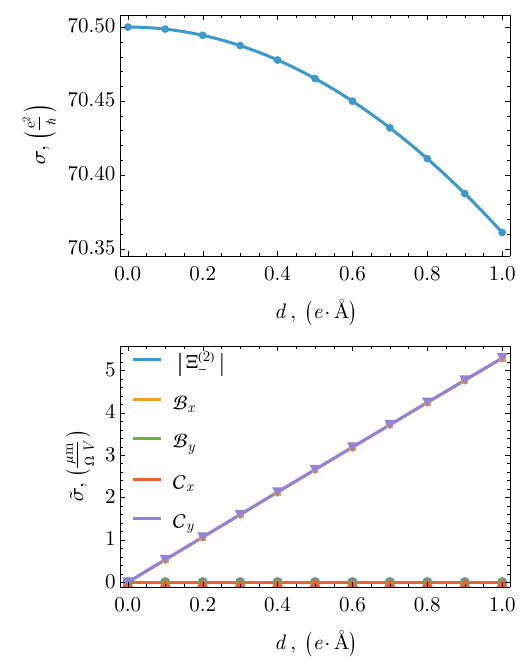}
\caption{ \add{(Top) Linear conductivity as a function of dipole moment. (Bottom) Components of the nonlinear conductivity tensor as a function of dipole moment, consistent with the symmetry analysis. Solid lines indicate the analytical solution using continued fractions (see SI and End Matter), while scatter points show numerical solution for angular harmonics obtained using Fast Fourier Transform. We used $\varepsilon_r = 8.5, m = 0.3 m_e, n_{imp} a^2 = 7.5 \times 10^{-5} \text{eV}^2 \text{\AA}^2, E_F = 6.6 \times 10^{-4} \text{eV}$, which describes ZnO case from Table \ref{tab:2DEG_params}, with $d_y=0$ for both panels. }  }
\label{fig2DEG}
\end{figure}

\textbf{Solutions for distribution functions, $\sigma$, and $\tilde{\sigma}$.}
Performing Fourier decompositions of $\Phi_{1}(\bk),\Phi_{2,1}(\bk)$ and $\Phi_{2,2}(\bk)$ 
\add{ as $\Phi_1 (\theta) = \sum_{n=1}^{\infty} \left( \Phi_1^{c,n} \cos n\theta + \Phi_1^{s,n} \sin n\theta \right),$ etc., } one can analytically solve for \emph{all} distribution functions and calculate $\sigma$, and $\tilde{\sigma}$. Introducing 
\begin{equation}
\begin{gathered}
    \alpha  = \hbar^4 \frac{e k_F \left( E_x + i E_y\right)}{m^2 n_{imp} a^2}, \;\;\;\;\;
    \beta = \frac{k_F e}{\hbar \add{\varepsilon_r}} \sqrt{\frac{m}{2 \mu}}, \\
    \bar{d} = d_x - i d_y, \;\;\;\;\; Z_n = \Phi_1^{c,n} + i \Phi_1^{s,n}, \\
    a_n =  \frac{-4(n+1)-2}{4(n+1)^2-1}, \;\;\;\;\;
    b_n = \frac{4(n-1)-2}{4(n-1)^2-1}
\end{gathered}
\end{equation}
one can cast the integral equation for $\Phi_1 (\bk)$ into a compact algebraic form 
\begin{eqnarray}
    Z_1 &=& \alpha + \beta a_1 \bar{d} Z_2, \\
    Z_n &=& \beta a_n \bar{d} Z_{n+1} + \beta b_n d Z_{n-1}.
\end{eqnarray}
This may be solved recursively 
\begin{equation}
    \begin{gathered}
        \rho_n = \frac{Z_{n+1}}{Z_n}, \;\;\;\;\;
        \rho_{n-1} = \frac{\beta b_n d}{1 - \beta a_n \bar{d} \rho_n},
    \end{gathered}
\end{equation}
leading to a continued fraction representation for $\rho_1$,
\begin{equation}
    \rho_1 = \frac{\beta b_2 d}{1 - \frac{\beta a_2 \bar{d} \beta b_3 d}{1 - \frac{\beta a_3 \bar{d} \beta b_4 d}{...}}}.
    \label{rho1_cont_frac}
\end{equation}
Eq. \eqref{rho1_cont_frac} can be evaluated numerically with arbitrary precision. Analogous closed-form expressions for $\Phi_{2,1} (\bk)$ and $\Phi_{2,2}(\bk)$ are presented in the SI. The continued fraction structure of the solutions is indicative of the coupling between different angular harmonics, resembling the results of \cite{Kryhin2,NazaryanLevitov}.
The distribution function solutions allow for a direct evaluation of the linear 
\begin{equation}
    j_{\gamma}^{(1)} = \frac{e}{\hbar} \frac{k_F}{4 \pi} \left( \mathrm{Re} \left( \frac{\alpha}{1- \beta a_1 \bar{d} \rho_1} \right), \mathrm{Im} \left( \frac{\alpha}{1- \beta a_1 \bar{d} \rho_1} \right) \right)_{\gamma},
    \label{lin_cond}
\end{equation}
and nonlinear current density \add{
\begin{widetext}
\begin{eqnarray}
    j_{\alpha} &=& \frac{\pi e k_F}{\hbar (2 \pi)^2} \left( \Phi_{2,1}^{c,1}, \Phi_{2,1}^{s,1} \right)_{\alpha} -  \frac{e}{\hbar} \frac{3 m}{4 \pi k_F \hbar^2} 
    \beta a_1 \left(  \mathrm{Re} \left( \bar{d} N_2 \right), \mathrm{Im} \left( \bar{d} N_2 \right) \right)_{\alpha} - \\
    &-&  \frac{e^2}{\hbar} \frac{3 \beta E_{\gamma}}{8 \pi \hbar g} 
    \begin{pmatrix}
        \left( a_2 \mathrm{Re} \left( \bar{d} Z_3 \right) + b_2 \mathrm{Re} \left( d Z_1 \right) \right) && \left( a_2 \mathrm{Im} \left( \bar{d} Z_3 \right) + b_2 \mathrm{Im} \left( d Z_1 \right) \right) \\
        \left( a_2 \mathrm{Im} \left( \bar{d} Z_3 \right) + b_2 \mathrm{Im} \left( d Z_1 \right) \right) && -\left( a_2 \mathrm{Re} \left( \bar{d} Z_3 \right) + b_2 \mathrm{Re} \left( d Z_1 \right) \right)
    \end{pmatrix}_{\alpha \gamma}
\end{eqnarray}
\end{widetext} 
}
In Eq. \eqref{lin_cond}, $\sigma \propto |\bd|^2$ as $\rho_1$ is a function of $d$, which is in accordance with symmetry analysis, Eq. \eqref{P_phenom_correct_no_Pz}.

\bibliography{biblio}

\end{document}